\documentclass{iaus}
\usepackage{graphicx}

\title[NStED-Stellar] %% give here short title %%
{The NStED Stellar and Exoplanet Hosting Star Service}

\author[S. Ramirez et al.]   %% give here short author list %%
{S. Ramirez$^{2,4,5}$,
B. Ali$^{2,4}$,
R. Baker$^{2,4}$,
G.B. Berriman$^{1,2,4}$,
K. von Braun$^{1,4}$,
N-M. Chiu$^{2,4}$,
D.R. Ciardi$^{1,4}$,
J. Good$^{2,4}$,
S.R. Kane$^{1,4}$,
A.C. Laity$^{2,4}$,
D.L. McElroy$^{2,4}$,
S. Monkewitz$^{2,4}$,
A.N. Payne$^{1,2,4}$,
M. Schmitz$^{2,4}$,
J.R. Stauffer$^{3,4}$,
P.L. Wyatt$^{1,2,4}$,
\and A. Zhang$^{2,4}$}

\affiliation{$^1$ Michelson Science Center;
$^2$ Infrared Processing and Analysis Center;
$^3$ Spitzer Science Center;
$^4$ California Institute of Technology;
$^5$ email: {\tt solange@ipac.caltech.edu}}

\pubyear{2008}
\volume{253}  %% insert here IAU Symposium No.
\pagerange{1--4}
\jname{Transiting Exoplanets}
\editors{Editor, eds.}
\begin{document}

\maketitle

%%%%%%%%%%%%%%%%%%%%%%%%%%%%%%%%%%%%%%%%%%%%%%%%%%%%%%%%%%%%%%%%%%%%

\begin{abstract}
The NASA Star and Exoplanet Database (NStED) is a general purpose
stellar archive with the aim of providing support for NASA's planet
finding and characterization goals, stellar astrophysics, and the
planning of NASA and other space missions. There are two principal
components of NStED: a database of (currently) 140,000 nearby stars
and exoplanet-hosting stars, and an archive dedicated to high
precision photometric surveys for transiting exoplanets. We present a
summary of the NStED stellar database, functionality, tools, and user
interface. NStED currently serves the following kinds of data for
140,000 stars (where available): coordinates, multiplicity, proper
motion, parallax, spectral type, multiband photometry, radial
velocity, metallicity, chromospheric and coronal activity index, and
rotation velocity/period.  Furthermore, the following derived
quantities are given wherever possible: distance, effective
temperature, mass, radius, luminosity, space motions, and
physical/angular dimensions of habitable zone. Queries to NStED can be
made using constraints on any combination of the above parameters. In
addition, NStED provides tools to derive specific inferred quantities
for the stars in the database, cross-referenced with available
extra-solar planetary data for those host stars.  NStED can be
accessed at \tt{http://nsted.ipac.caltech.edu}.

\keywords{astronomical data bases: miscellaneous; catalogs; surveys;
  time; stars: variables; planetary systems}
\end{abstract}

%%%%%%%%%%%%%%%%%%%%%%%%%%%%%%%%%%%%%%%%%%%%%%%%%%%%%%%%%%%%%%%%%%%%

\firstsection % if your document starts with a section,

%%%%%%%%%%%%%%%%%%%%%%%%%%%%%%%%%%%%%%%%%%%%%%%%%%%%%%%%%%%%%%%%%%%%

\section{The NStED Services}

The NASA Star and Exoplanet Database (NStED) is dedicated to
collecting and serving vital published data involved in the search
for and study of extrasolar planets and their host stars. The stellar
and exoplanet services provide access to stellar parameters of
potential exoplanet bearing stars along with exoplanet parameters. The
stellar services provided by NStED include the following:
\begin{itemize}
\item Data related to relatively bright nearby stars.
\item The ability to query for individual stars or search by
  stellar/planetary parameters.
\item Published images, spectra, and time series data related to the
  stars in the database.
\end{itemize}
Complementary to this are the exoplanet services, which includes the
following:
\begin{itemize}
\item General data and pubished parameters for the known exoplanets
  and host stars.
\item Photometric and radial velocity data related to the known
  exoplanets.
\item A dedicated interface related to exoplanet transit surveys (see
  companion paper by von Braun et al. (2008) in these proceedings).
\end{itemize}
Figure 1 shows an example plot produced using the data served by NStED.

\begin{figure}
  \begin{center}
    \includegraphics[width=\textwidth]{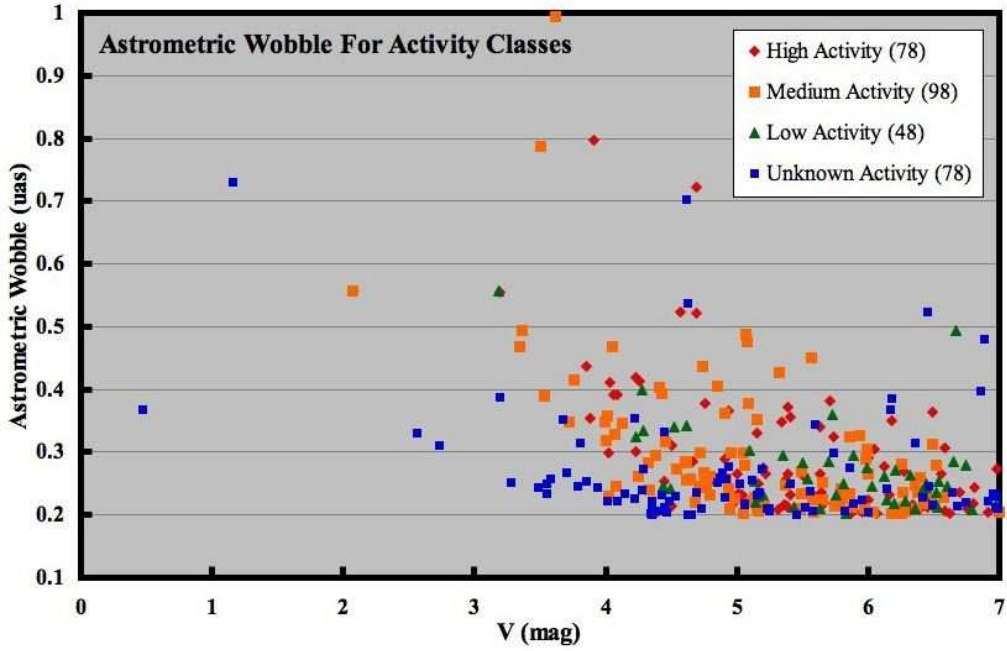}
    \caption{Plot of predicted astrometric wobble for an Earth-sized
      planet in the habitable zone vs. the apparent V magnitude of the
      stars, generated using data served by NStED. The stars are
      sorted by activity level estimates from the R'(HK) index, S
      index, and X-ray luminosity.}
  \end{center}
\end{figure}

%%%%%%%%%%%%%%%%%%%%%%%%%%%%%%%%%%%%%%%%%%%%%%%%%%%%%%%%%%%%%%%%%%%%

\section{Stellar Content for NStED}

NStED's stellar and exoplanet content is composed of published tabular
data, derived and calculated quantities, and associated data including
images, spectra, and time series. An example spectra from the N2K
consortium (\cite{fis05}) contained within NStED is shown in Figure 2.
NStED's core set of stars is derived from the Hipparcos,
Gliese-Jahreiss, and Washington Double Star catalogs. The total number
of Hipparcos and Gliese-Jahreiss stars within NStED is approximately
140,000. A summary of the stellar parameters and data within NStED is
shown in Table 1. NStED currently supports complex multi-faceted
queries on approximately 75 astrophysical stellar and exoplanet
parameters.

\begin{figure}
  \begin{center}
    \includegraphics[width=\textwidth]{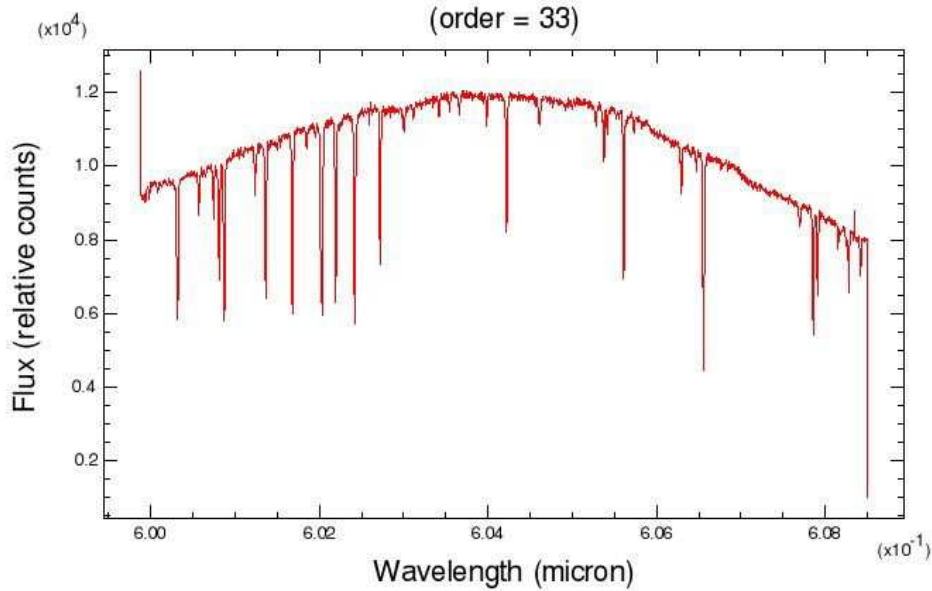}
    \caption{N2K spectrum of HD 804.}
  \end{center}
\end{figure}

\begin{table}
  \begin{center}
    \caption{Summary of stellar content within NStED.}
    \begin{tabular}{@{}lll}
      \hline
      Published Parameters & Derived Parameters & Associated Data \\
      \hline
      Position, Distances & Temperature      & Images  \\
      Kinematics          & Luminosity       & Spectra \\
      Photometry, Colors  & Radius           & \\
      Spectral Type       & Mass             & \\
      Luminosity Class    & LSR Space Motion & \\
      Metallicity         & & \\
      Rotation            & & \\
      Activity Indicators & & \\
      Variability         & & \\
      Multiplicity        & & \\
      \hline
    \end{tabular}
  \end{center}
\end{table}

%%%%%%%%%%%%%%%%%%%%%%%%%%%%%%%%%%%%%%%%%%%%%%%%%%%%%%%%%%%%%%%%%%%%

\section{Exoplanet Content for NStED}

In order to facilitate future exoplanet studies, NStED maintains an
up-to-date list of exoplanetary systems and associated stellar data by
monitoring daily the literature and making weekly updates to the
database. These data include high-precision lightcurves, such as the
lightcurve for the transiting planet TrES-2 shown in Figure 3. The
predicted signatures of exoplanets are also calculated to aid users in
selection of stars appropriate for planet searching and
characterization. The exoplanet signature predictions include
habitable zone sizes, astrometric and radial velocity wobbles, and
transit depths. A summary of the exoplanet parameters and data within
NStED is shown in Table 2.

\begin{figure}
  \begin{center}
    \includegraphics[width=\textwidth]{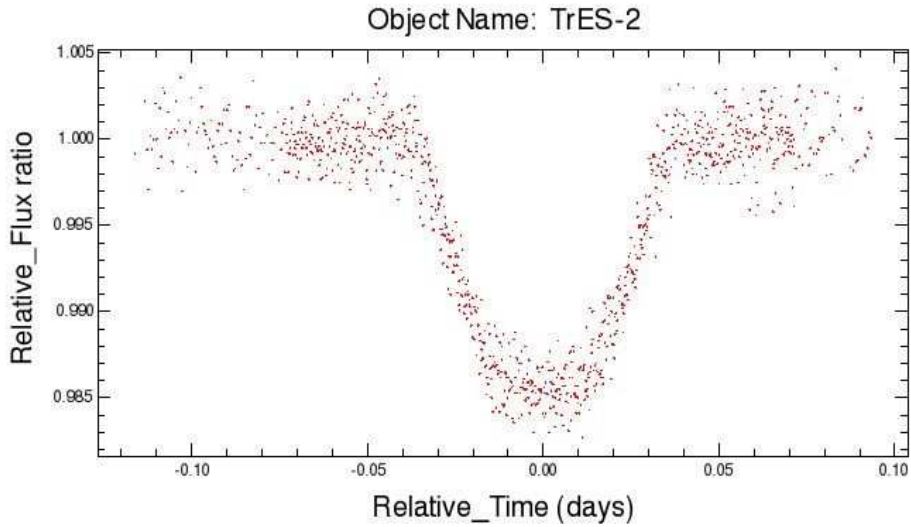}
    \caption{Lightcurve of the transiting exoplanet TrES-2
      (\cite{odo06}).}
  \end{center}
\end{figure}

\begin{table}
  \begin{center}
    \caption{Summary of exoplanet content within NStED.}
    \begin{tabular}{@{}lll}
      \hline
      Published Parameters & Predicted Parameters & Associated Data \\
      \hline
      Number of Planets & Habitable Zone          & High Contrast Images \\
      Planetary Mass    & Astrometric Wobble      & Lightcurves  \\
      Orbital Period    & Radial Velocity Wobble  & \\
      Orbital semi-major axis & Earth V Magnitude & \\
      Orbital Eccentricity & Earth 10 $\mu$m flux density & \\
      Link to entry in the & & \\
      Exoplanet Encyclopaedia & & \\
      \hline
    \end{tabular}
  \end{center}
\end{table}

%%%%%%%%%%%%%%%%%%%%%%%%%%%%%%%%%%%%%%%%%%%%%%%%%%%%%%%%%%%%%%%%%%%%

\section{Summary}

The NStED Stellar and Exoplanet Hosting Star Service provides access
to data relevant to exoplanet host stars and bright stars from major
catalogues. The search query tools and cross-referencing capabilities
make this a powerful engine through which to aid in exoplanet survey
programs. NStED provides not only tabular data, but also a wide
variety of associated data including images, spectra, and time series
(radial velocity and photometric) obervations. NStED is continually
updated to reflect the latest results in the literature and to provide
published data access to the broader astronomical community. NStED is
accessible via {\tt{http://nsted.ipac.caltech.edu}}.

\begin{figure}
  \begin{center}
    \includegraphics[width=\textwidth]{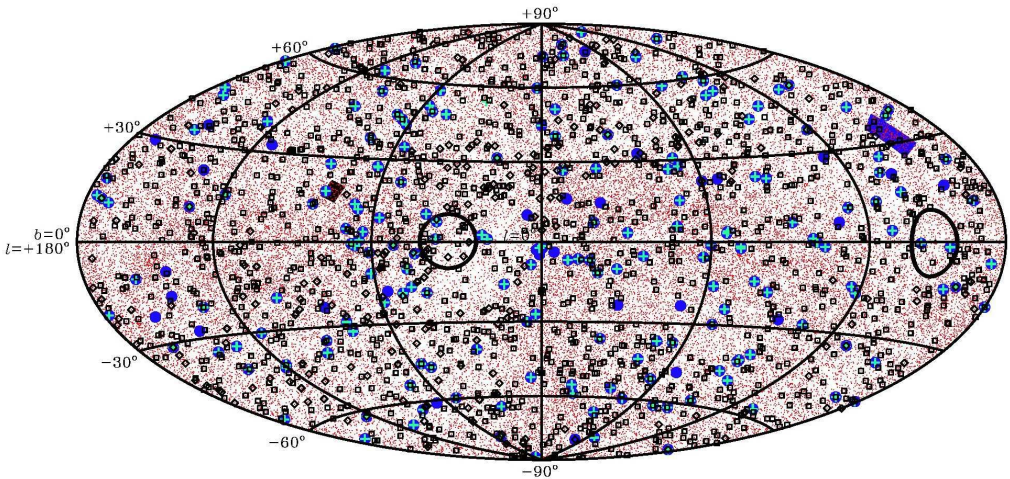} 
    \caption{Aitoff projection of the contents of NStED. Small dots:
      dwarf stars (for clarity, the giant stars are not plotted);
      large dots: exoplanet hosting stars; large plus signs: stars
      with radial velocity curves or photometric lightcurves; open
      squares/diamonds: stars with images/spectra. For an explanation
      of the remaining features, see companion paper on the NStED
      Exoplanet Transit Survey Service by von Braun et al. (2008) in
      these proceedings.}
  \end{center}
\end{figure}

%%%%%%%%%%%%%%%%%%%%%%%%%%%%%%%%%%%%%%%%%%%%%%%%%%%%%%%%%%%%%%%%%%%%

%%%%%%%%%%%%%%%%%%%%%%%%%%%%%%%%%%%%%%%%%%%%%%%%%%%%%%%%%%%%%%%%%%%%

\end{document}